\documentclass{article}
\usepackage{spconf,amsmath,graphicx,hyperref}
\hypersetup{hidelinks}
\usepackage{amssymb}
\title{BAMU: Bitstream-Aware Marginal-Utility Allocation for Frozen Pretrained Neural Speech Codecs}
\name{
\begin{tabular}{c}
Mingyu Zhao$^{1,*}$,
Zijian Lin$^{1,*}$,
Yutang Feng$^{2}$,
Jiatao Chen$^{2}$,
Fan Wang$^{1}$,
Jiehui Luo$^{3}$,\\
Yuhao Ding$^{3}$,
Jinchao Zhang$^{2,\dagger}$,
Zhiyong Wu$^{1,\dagger}$
\end{tabular}
\thanks{$^{*}$Equal contribution.
$^{\dagger}$Corresponding authors.}
}

\address{
$^{1}$Tsinghua Shenzhen International Graduate School,
Tsinghua University, Shenzhen, China\\
$^{2}$Tencent, Shenzhen, China\\
$^{3}$Central Conservatory of Music, Beijing, China\\
\texttt{\{zmy24,linzj24,wang-f25\}@mails.tsinghua.edu.cn}\\
\texttt{\{yutangfeng,wietchen,dayerzhang\}@tencent.com}\\
\texttt{\{luojiehui,dingyuhao\}@mail.ccom.edu.cn}
\texttt{\quad zywu@sz.tsinghua.edu.cn}
}
\begin{document}

%\ninept
%
\maketitle
\begin{abstract}
Pretrained neural speech codecs typically use a fixed residual vector
quantization (RVQ) depth for all frames, ignoring temporal variation in
quantization difficulty. We propose BAMU, a bitstream-aware dynamic RVQ
allocation framework for frozen pretrained codecs. A lightweight,
rate-independent predictor estimates frame- and layer-wise marginal
latent-distortion reductions, while a constrained allocator selects
prefix-valid depths under an exact serialized-size budget. Experiments on EnCodec
and DAC over LibriSpeech, together with VCTK evaluation, show consistent
EnCodec gains and DAC improvements mainly at medium and high rates. A 30-listener study confirms a MOS improvement from 3.449 to 3.780 over
matched fixed-depth coding.
\end{abstract}

\section{Introduction}
\label{sec:introduction}

Neural speech codecs commonly combine learned encoder--decoder
architectures with residual vector quantization (RVQ) for high-quality
compression at low bitrates
\cite{gray1984residual,zeghidour2022soundstream,defossez2023encodec,
kumar2023dac,wu2023audiodec,du2024funcodec}. Conventional codecs typically apply the same RVQ prefix depth to every
frame, ignoring temporal variation and potentially wasting bits on frames
that benefit little from additional refinement.

Adaptive neural coding has been explored through quantizer dropout,
frame-wise importance prediction, variable temporal resolution, semantic
priors, and test-time code search
\cite{zeghidour2022soundstream,chae2025vrvq,zhang2025tfc,
li2026flexicodec,zhao2026spgcodec,
kim2025testtime}. VRVQ is most closely related to our work, as it varies
the number of active RVQ codebooks using predicted frame-wise importance
\cite{chae2025vrvq}. Unlike VRVQ, which learns variable-rate coding jointly with the codec,
BAMU performs post-hoc allocation on frozen pretrained codecs and
explicitly enforces the realized serialized-container budget. Other methods adapt temporal resolution
\cite{zhang2025tfc,li2026flexicodec}, introduce semantic guidance
\cite{zhao2026spgcodec}, or improve code selection at a prescribed
depth \cite{kim2025testtime}. This leaves a practical gap in post-hoc
frame-wise RVQ allocation for frozen pretrained codecs, particularly
when the realized container size rather than a nominal rate must be
enforced.

\begin{figure*}[t]
    \centering
    \includegraphics[width=0.9\textwidth]{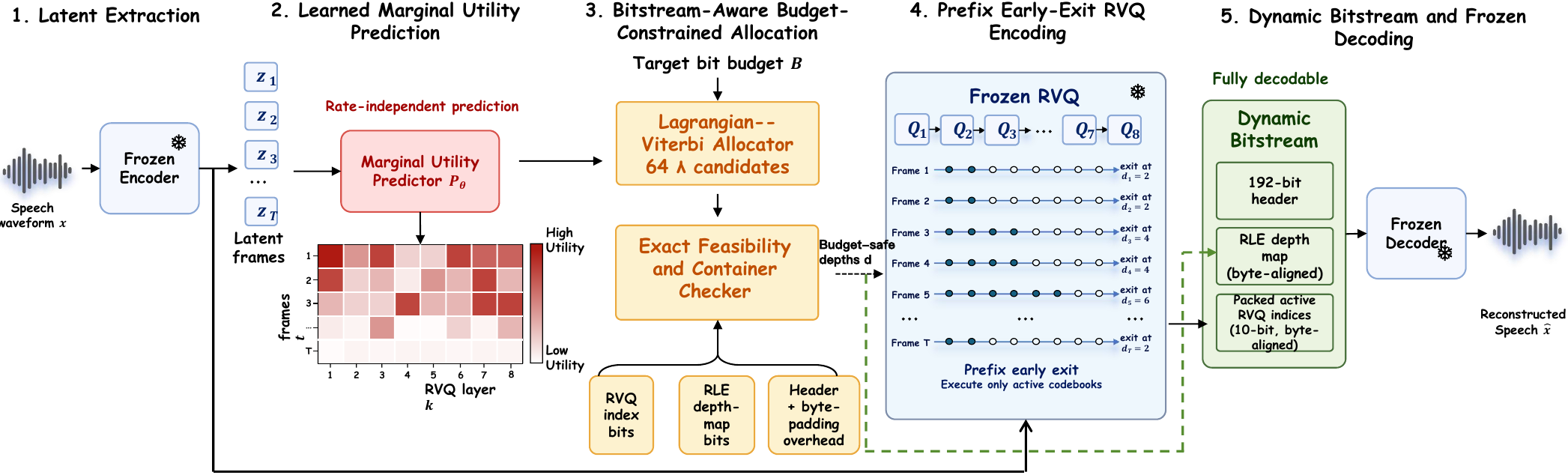}
    \caption{Overview of the proposed framework. A lightweight predictor
    estimates frame- and layer-wise marginal utilities, and a
    bitstream-aware allocator selects prefix-valid RVQ depths under a
    realized container budget. The pretrained codec remains frozen.}
    \label{fig:framework}
\end{figure*}

We propose a bitstream-aware dynamic RVQ allocation framework for frozen
pretrained neural speech codecs, as shown in Fig.~\ref{fig:framework}.
A lightweight predictor estimates the marginal latent-distortion
reduction contributed by each RVQ layer at every frame, independently of
the target bitrate. A constrained allocator then selects prefix-valid
depths and verifies the exact serialized size, including active indices,
the run-length-coded depth map, the header, and byte padding. The
allocation is realized through prefix early exit and a fully decodable
dynamic bitstream, while the original codec remains frozen.

Our contributions are threefold. First, we introduce rate-independent,
frame- and layer-wise marginal-utility prediction for multiple operating
rates. Second, we develop a bitstream-aware allocator with exact
container-budget enforcement, prefix early exit, and complete decoding.
Third, extensive experiments demonstrate consistent EnCodec gains and
DAC improvements primarily at medium and high rates under matched
realized budgets.

\section{Proposed Method}
\label{sec:method}

\subsection{Marginal-Utility Prediction}
\label{subsec:predictor}

Let the frozen encoder map waveform $x$ to latent frames
$\mathbf{z}=E(x)=\{z_t\}_{t=1}^{T}$, where
$z_t\in\mathbb{R}^{C}$. For an RVQ with $K$ codebooks, let
$r_{t,0}=z_t$ and let $r_{t,k}$ denote the residual after the first
$k$ codebooks. We define the channel-normalized distortion and marginal
utility as
\begin{equation}
D_{t,k}=\frac{1}{C}\|r_{t,k}\|_2^2,\qquad
\Delta D_{t,k}=[D_{t,k-1}-D_{t,k}]_{+},
\label{eq:marginal_utility}
\end{equation}
where $[a]_{+}=\max(a,0)$. Negative reductions are clipped to zero in the main system;
Sec.~\ref{subsec:analysis} additionally evaluates signed marginal targets.

Rather than predicting a bitrate-specific frame depth, a lightweight
predictor $P_{\theta}$ estimates all frame- and layer-wise utilities.
To stabilize regression, the targets are transformed as
\begin{equation}
y_{t,k}=\log\left(1+\frac{\Delta D_{t,k}}{s}\right),
\qquad s=10^{-4}.
\end{equation}
The predictor is trained with a masked Smooth-L1 loss,
\begin{equation}
\mathcal{L}_{\mathrm{utility}}
=
\frac{1}{|\Omega|}
\sum_{(t,k)\in\Omega}
\operatorname{SmoothL1}
(\widehat{y}_{t,k},y_{t,k}),
\label{eq:utility_loss}
\end{equation}
where $\Omega$ excludes padded frames. At inference,
\begin{equation}
\widehat{\Delta D}_{t,k}
=
s[\exp(\widehat{y}_{t,k})-1]_{+}.
\end{equation}
The predictor takes frozen encoder latents as input and consists of
three non-causal temporal Conv1d blocks with kernel sizes 5, 3, and 3,
followed by a $1\times1$ output layer. The hidden dimensions are 128,
128, and 64, with GELU activation and GroupNorm.

\subsection{Bitstream-Aware Depth Allocation}
\label{subsec:allocator}

Each frame is assigned a prefix-valid depth
$d_t\in\{1,\ldots,K\}$. The predicted utility of an allocation
$\mathbf{d}$ is
\begin{equation}
U(\mathbf{d})
=
\sum_{t=1}^{T}\sum_{k=1}^{d_t}
\widehat{\Delta D}_{t,k}.
\end{equation}
We seek
\begin{equation}
\max_{\mathbf{d}}\;U(\mathbf{d})
\quad\mathrm{s.t.}\quad
R_{\mathrm{actual}}(\mathbf{d})\leq B,
\label{eq:allocation}
\end{equation}
where $B$ is the target container budget. Adjacent frames may be grouped
into blocks sharing one depth; the default block size is four.
For block $\mathcal{B}_n$, define its cumulative utility at depth $k$ as
\begin{equation}
U_{n,k}
=
\sum_{t\in\mathcal{B}_n}\sum_{j=1}^{k}
\widehat{\Delta D}_{t,j}.
\end{equation}

We employ parallel Lagrangian relaxation with GPU-batched Viterbi
decoding. For 64 multiplier candidates, including $\lambda=0$ and 63
logarithmically spaced positive values, the allocator optimizes
\begin{equation}
\sum_n U_{n,d_n}
-
\lambda\left(
10\sum_n L_n d_n
+
\beta\sum_{n>1}\mathbf{1}[d_n\neq d_{n-1}]
\right),
\label{eq:lagrangian}
\end{equation}
where $n$ indexes blocks, $L_n$ is the number of frames in block $n$,
and each RVQ index occupies 10 bits. The transition coefficient
$\beta=6$ approximates depth-map cost during optimization but is not an
actual bitstream field.

After backtracking, each candidate is evaluated using its exact raw
payload, where $R_{\mathrm{code}}^{\mathrm{raw}}=10\sum_t d_t$.
For the $K=8$ setting used in our experiments, each depth-map run stores
a 3-bit depth value and an Elias-gamma-coded run length~\cite{elias1975universal}:
\begin{equation}
R_{\mathrm{depth}}^{\mathrm{raw}}
=
\sum_m
\left(
3+2\lfloor\log_2\ell_m\rfloor+1
\right).
\label{eq:rle_bits}
\end{equation}

The fixed 192-bit header is shared by the dynamic and matched fixed-depth
containers and is therefore omitted during allocation. Let
$B_{\mathrm{payload}}^{\mathrm{fixed}}$ denote the raw code-plus-depth
payload of the matched fixed-depth stream. Because the code and depth
payloads are independently byte aligned, we conservatively reserve
14 bits for their worst-case combined padding. Only candidates satisfying
\begin{equation}
R_{\mathrm{code}}^{\mathrm{raw}}
+
R_{\mathrm{depth}}^{\mathrm{raw}}
\leq
B_{\mathrm{payload}}^{\mathrm{fixed}}-14
\label{eq:reserved_budget}
\end{equation}
are retained.

The feasible candidate with the highest predicted utility is selected.
A heap-based greedy step then proposes prefix-valid whole-block layer
additions in descending marginal-utility-per-code-bit order. Each proposal
is accepted only after recomputing the exact raw RLE payload.

Finally, an external container-level checker computes the realized
dynamic-stream size as
\begin{equation}
R_{\mathrm{actual}}^{\mathrm{dyn}}(\mathbf{d})
=
192
+
8\left\lceil
\frac{R_{\mathrm{depth}}^{\mathrm{raw}}(\mathbf{d})}{8}
\right\rceil
+
8\left\lceil
\frac{R_{\mathrm{code}}^{\mathrm{raw}}(\mathbf{d})}{8}
\right\rceil ,
\label{eq:actual_bits}
\end{equation}
where the two payloads are independently byte aligned. The same
serializer is used to obtain the realized size
$R_{\mathrm{actual}}^{\mathrm{fixed}}$ of the matched fixed-depth
container, and the final allocation is accepted only if
\begin{equation}
R_{\mathrm{actual}}^{\mathrm{dyn}}(\mathbf{d})
\leq
R_{\mathrm{actual}}^{\mathrm{fixed}}.
\label{eq:matched_container_budget}
\end{equation}

\subsection{Prefix Early Exit and Dynamic Bitstream}
\label{subsec:bitstream}

The selected allocation is realized through prefix early exit. At RVQ
stage $k$, only frames satisfying $d_t\geq k$ are processed by codebook
$Q_k$; frames that have reached their assigned depths skip subsequent
codebook searches. This is numerically equivalent to full RVQ evaluation
followed by masking inactive indices.

Both fixed and dynamic streams use the same explicitly specified
prototype serializer. A fixed-depth stream represents its constant depth using a single RLE
run. Each stream stores a 192-bit header, the run-length-coded
depth map, and the packed active RVQ indices. During decoding, the depth map
specifies the number of prefix indices associated with each frame, and
the reconstructed latent sequence is passed to the frozen decoder $G$:
\begin{equation}
\widehat{x}
=
G\left(
\left\{
\sum_{k=1}^{d_t}q_{t,k}
\right\}_{t=1}^{T}
\right).
\end{equation}

\section{Experiments and Results}
\label{sec:experiments}

\subsection{Setup}
\label{subsec:setup}

We evaluate two pretrained 24-kHz neural codecs, EnCodec
\cite{defossez2023encodec} and DAC~\cite{kumar2023dac}, using the first
eight RVQ codebooks of each codec. The original codecs remain frozen,
and a separate utility predictor is trained for each. Both codecs use 1024-entry codebooks, so each RVQ index is represented
by 10 bits.

Training uses LibriSpeech \texttt{train-clean-100}
\cite{panayotov2015librispeech}, comprising 28,539 utterances,
100.6 hours, and approximately 27.2 million latent frames;
\texttt{dev-clean} is used for validation. Evaluation is conducted on
\texttt{test-clean} and \texttt{test-other}, containing 2,620 and
2,939 utterances, respectively. We additionally evaluate the EnCodec
predictor on 1,100 VCTK utterances
\cite{yamagishi2019vctk} without VCTK training or fine-tuning.

The EnCodec and DAC predictors contain 157,128 and 730,568 trainable
parameters, respectively. Both are trained for 40 epochs using AdamW with an initial learning
rate of $3\times10^{-4}$, cosine decay to $10^{-6}$, and 512-frame
crops. We use weight decay $10^{-4}$, gradient clipping at 5.0, and
per-GPU batch sizes of 32 and 16 for EnCodec and DAC, respectively. Unless otherwise stated, allocation
uses four-frame blocks, $\beta=6$, and 64 multiplier candidates.

Fixed-depth RVQ is the primary baseline. EnCodec dynamic points are
matched to fixed depths two through seven, whereas DAC is evaluated at
depths three through five. For EnCodec, additional comparisons at representative operating points
include random, periodic, waveform-energy, and ground-truth
residual-utility allocation.
We report PESQ~\cite{rix2001pesq}, STOI~\cite{taal2011stoi}, and
ViSQOL~\cite{chinen2020visqol}. All main fixed-versus-dynamic
comparisons use exact serialized-container accounting rather than
nominal average depth.

\begin{figure}[t]
    \centering
    \includegraphics[width=0.85\columnwidth]{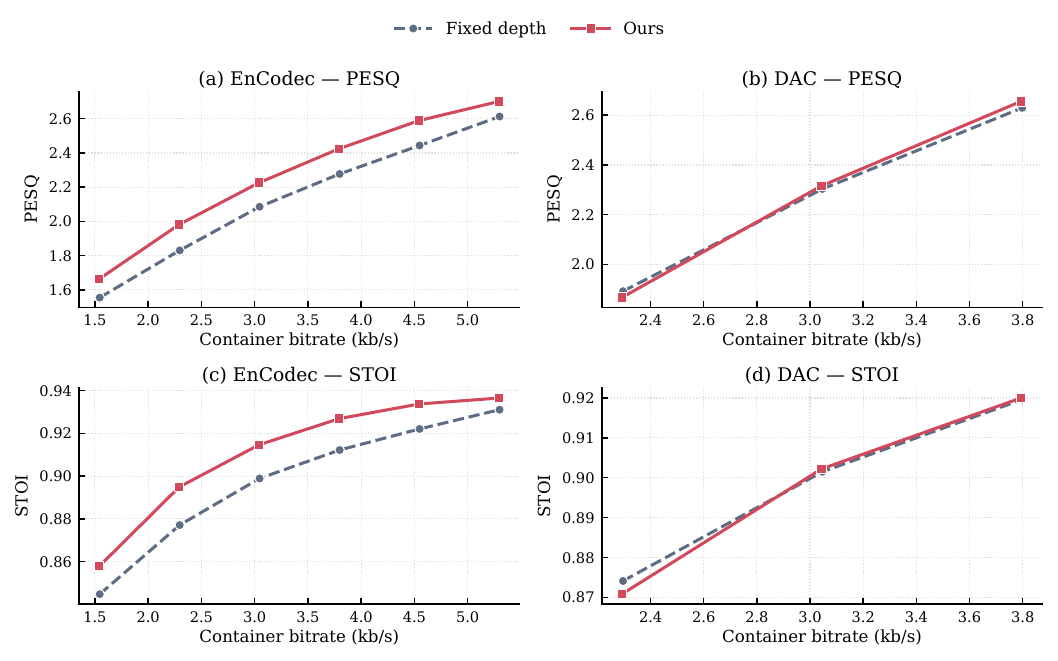}
    \caption{Rate--distortion performance on LibriSpeech
    \texttt{test-clean} using realized container bitrates. Each dynamic
    point does not exceed its matched fixed-depth container size.}
    \label{fig:rd}
\end{figure}

\subsection{Rate--Distortion Performance}
\label{subsec:rd}

Fig.~\ref{fig:rd} shows that dynamic allocation consistently improves
EnCodec. On \texttt{test-clean}, PESQ gains range from 0.088 to 0.151
at the matched depth-two to depth-seven budgets; at depth four, PESQ
increases from 2.086 to 2.226. The corresponding gains on
\texttt{test-other} are 0.086--0.149, and STOI improves at every
matched operating point. All EnCodec PESQ bootstrap confidence intervals
exclude zero, and both paired $t$-tests and Wilcoxon signed-rank tests
are significant. At depth four on \texttt{test-clean}, Cohen's
$d_z$ is 1.12 and the utterance-level win rate is 89.6\%.

DAC exhibits a stronger rate dependence. On \texttt{test-clean}, PESQ
changes by $-0.0256$, $+0.0123$, and $+0.0259$ at depths three, four,
and five, respectively. On \texttt{test-other}, the corresponding
changes are $+0.0078$, $+0.0412$, and $+0.0460$. The lowest-rate
\texttt{test-other} result is conservatively treated as approximately
matched because its Wilcoxon test is not significant. Thus, DAC benefits
most consistently at medium and high rates.

ViSQOL shows the same overall trend, with EnCodec generally improving
and DAC benefiting mainly at medium and high rates.

Without VCTK adaptation, EnCodec improves PESQ from 2.303 to 2.490 at
the matched depth-four budget, indicating cross-dataset generalization.

\subsection{Baselines and Ablations}
\label{subsec:analysis}

\begin{table}[t]
\centering
\caption{EnCodec PESQ on LibriSpeech \texttt{test-clean}. Auxiliary
baselines use raw-budget matching; BAMU additionally enforces the packed
container constraint. GT utility is a privileged frame-level reference.}
\label{tab:baseline}
\begin{tabular}{lcc}
\hline
Method & Depth-3 & Depth-4\\
\hline
Fixed depth &1.831&2.086\\
Random&1.438&1.581\\
Periodic&1.491&1.646\\
Waveform energy&1.751&2.006\\
BAMU&1.982&2.226\\
GT utility&2.041&2.284\\
\hline
\end{tabular}
\end{table}
Table~\ref{tab:baseline} shows that BAMU outperforms random, periodic,
and waveform-energy allocation. GT utility is a privileged frame-level
reference requiring all candidate-layer residuals. For DAC, predicted and ground-truth utilities are evaluated
under the same block-wise protocol. At depth three, even ground-truth
allocation reaches only 1.8779 PESQ, below fixed-depth coding at 1.8922.
Negative gains occur in 12.83\% of DAC frame--layer pairs, although their
total magnitude is only 2.07\% of the positive-gain magnitude. For the signed variant, we use
$\delta D_{t,k}=D_{t,k-1}-D_{t,k}$ and
$y_{t,k}=\operatorname{sgn}(\delta D_{t,k})
\log(1+|\delta D_{t,k}|/s)$. Replacing
clipped targets with signed marginal targets improves depth-three PESQ
from 1.8666 to 1.8810 on \texttt{test-clean} and from 1.8645 to 1.8759
on \texttt{test-other}, but slightly reduces the depth-four and
depth-five results. Moreover, signed depth-three allocation remains below
fixed-depth coding on \texttt{test-clean} (1.8810 versus 1.8922).
Thus, clipping contributes to but does not fully explain the lowest-rate
degradation, indicating a limitation of latent-MSE utility as a perceptual
allocation objective.

Computed separately within each RVQ layer, Pearson correlations range
from 0.908 to 0.952 and Spearman correlations from 0.945 to 0.960, with
a mean top-quartile overlap of 0.822. Three training seeds produce less than 0.0011 PESQ
variation at representative operating points.

At depths three and four, the main temporal marginal-utility predictor
obtains PESQ values of 1.9822 and 2.2265. Pointwise and reduced-width
predictors perform similarly, whereas cumulative-utility prediction
drops to 1.9261 and 2.1699, confirming the importance of marginal
targets. Switch penalties of zero, six, and twelve yield similar quality,
while larger values reduce depth-map variation. Block sizes of one, two,
four, and eight obtain depth-four PESQ values of 2.260, 2.249, 2.226,
and 2.186, respectively, compared with 2.086 for fixed depth four.
Block size four is used as a practical quality--efficiency compromise.

The GPU-batched allocator causes no budget violations and retains more
than 99\% of the utility achieved by the CPU reference allocator.

\subsection{Bitstream, Efficiency, and Subjective Quality}
\label{subsec:system}

All main comparisons use exact container-size accounting. We additionally
materialize and decode dynamic bitstreams for 200 DAC
\texttt{test-clean} utterances at depths three through five. All depth
maps and active indices are recovered exactly, with no failures and zero
maximum latent error. The corresponding realized dynamic bitrates are
2.290, 3.041, and 3.792~kb/s. Prefix early exit is also numerically
equivalent to full RVQ evaluation followed by masking.

On a single NVIDIA A100 GPU, for a representative ten-second DAC input,
the fixed and dynamic inference pipelines achieve RTFs of approximately
0.009 and 0.015, respectively, excluding audio I/O. Prefix early exit
adds only 2.1--2.5\% overhead to the codec hot path, while the
utterance-level allocator increases end-to-end inference latency by
approximately 63--70\%, representing the main computational overhead
of BAMU.

Finally, we conduct a randomized, anonymized P.800-style listening test
\cite{itu1996p800} with 25 utterances, four conditions, and 30 listeners,
yielding 3,000 valid ratings. Hidden reference, fixed depth eight, our
method, and fixed depth four obtain MOS values of 4.509, 4.040, 3.780,
and 3.449, respectively. Under the matched depth-four container budget,
our method improves MOS by 0.331, with a listener-level 95\% confidence
interval of $[0.247,0.414]$ and significance under both paired tests
($p<10^{-5}$).

\section{Conclusions}
\label{sec:conclusion}

We proposed BAMU, a bitstream-aware dynamic RVQ allocation framework for
frozen pretrained neural speech codecs. BAMU combines marginal-utility
prediction, exact container-budget enforcement, prefix early exit, and a
fully decodable dynamic bitstream. Experiments show consistent EnCodec
gains and DAC improvements primarily at medium and high rates, together
with exact bitstream recovery and significant subjective quality gains.

\noindent\textbf{Acknowledgment:}
This work is supported by National Natural Science Foundation of China
(62076144) and Shenzhen Science and Technology Program
(JCYJ20220818101014030).

\bibliographystyle{IEEEbib}
\bibliography{strings,refs}

\end{document}